\documentclass[preprint,12pt]{elsarticle}

\usepackage{amssymb}
\usepackage{amsmath}
\usepackage{textcomp}
\usepackage{xcolor}

\journal{IJPT}

\begin{document}

\begin{frontmatter}


\title{Machine Learning-based beam delivery time model for Mevion 250i with Hyperscan technology}

\author[maastro]{Giorgio Cartechini\corref{cor1}}
\ead{giorgio.cartechini@maastro.nl}
\cortext[cor1]{}
\author[unitn]{Francesco Giuseppe Cordoni} 
\author[maastro]{Mirko Unipan} 
\author[maastro]{Ilaria Rinaldi} 
\affiliation[maastro]{organization={Maastricht University Medical Centre+, Department of Radiation Oncology (Maastro), GROW School for Oncology and Reproduction},
             addressline={Doctor Tanslaan 12},
             city={Maastricht},
             postcode={6229 ET},
             country={The Netherlands}}

\affiliation[unitn]{organization={University of Trento, Department of Civil, environmental and mechanical engineering},
             addressline={via Mesiano, 77},
             city={Trento},
             postcode={38123},
             country={Italy}}

\begin{abstract}
\textbf{Purpose:} Accurate prediction of beam delivery time (BDT) is critical for operational efficiency, 4D dose calculations, and advanced proton therapy techniques. Despite its importance, no machine-specific BDT model exists for Mevion systems.

\textbf{Methods:} We developed the first machine learning–based model for the Mevion S250i Hyperscan system. Institutional machine log files from 11 patients (1120 machine log files) were used to extract features describing spot position, energy layer changes, Adaptive Aperture (AA) movements, and spot charge. Inter-pulse time ($\Delta T$) served as the target variable. A Random Forest model was trained with cross-validation and tested on held-out data. SHAP (SHapley Additive exPlanations) analysis quantified feature contributions.

\textbf{Results:} The model achieved mean absolute errors (MAE) ranging from 0.9 ms for short intervals ($<$50 ms) to 222 ms for long delays ($>$1000 ms). AA movements were the dominant global predictor for $\Delta T > 50$ ms, whereas spot positions and pulse charge dominated short intervals. Energy changes had a minor global influence but locally contributed to a large $\Delta T$, consistent with range modulator physics. The model was tested on two clinically relevant applications: volumetric repainting beam sequence and 4D dose recalculation for interplay evaluation. The predicted cumulative delivery times deviated by only -1.7\% from machine log files, and dosimetric metrics ($D_{98}$, $D_{95}$, and $V_{95}$) remained within intrinsic delivery variability.

\textbf{Conclusions:} This study presents the first machine learning–based BDT model for the Mevion S250i system, accurately capturing both predictive performance and machine-specific temporal dynamics. Explainable AI analysis using SHAP provided detailed insights into the operational characteristics of the system, highlighting the contributions of energy layer switching, Adaptive Aperture adjustments, and spot position shifts to delivery time. The proposed BDT model demonstrated strong predictive performance across the clinical applications evaluated, supporting its potential use for interplay assessment, 4D dose calculation, and delivery time–based plan optimization.

\end{abstract}




\end{frontmatter}



\section{Introduction}
\label{Intro}

Recent advancements in proton therapy have been driven by the need to reduce the size, weight, and cost of accelerator systems. Vendors have increasingly adopted superconducting magnet technology to achieve compact designs~\cite{henrotin2016commissioning}. This miniaturization not only lowers acquisition and operational costs but also enables integration into existing radiotherapy departments. An example is Mevion’s pencil beam scanning (PBS) system, which features a gantry-mounted superconducting synchrocyclotron equipped with a proton multi-leaf collimator and a nozzle-mounted range modulator \cite{vilches2020beam}. The Mevion S250i Hyperscan system is currently operational in the US, Asia, and Europe, in particular at the Maastro Proton Therapy in Maastricht, the Netherlands~\cite{vilches2020beam}, the first European facility. The latest development in this domain is the Mevion S250-FIT system, created in collaboration with Leo Cancer Care~\cite{volz2022considerations}. This system introduces a gantry-less PBS configuration with upright patient positioning, designed to fit within a standard LINAC vault~\cite{Mevion250FIT}. Such innovations are particularly well-suited for emerging techniques like proton arc therapy (PAT), which require dynamic and time-resolved beam delivery. In this system, beam energy selection is performed within the nozzle using a range modulator, resulting in larger spot sizes. To compensate, the Adaptive Aperture (AA) system, based on multi-leaf collimators, is employed to sharpen the lateral penumbra of each spot~\cite{silvus2024dosimetric, engwall2025shoot}. 

These technological innovations have a direct impact on the beam delivery characteristics of these systems. In scanned ion beam therapy, treatment planning involves defining energies, spot spatial coordinates, and particle weights \cite{janson2024treatment}. Therefore, understanding and predicting the beam delivery time (BDT) is crucial, not only for operational efficiency but also for clinical accuracy. In 4D dose calculations, BDT modeling enables accurate simulation of the interplay between tumor motion and spot delivery sequence~\cite{li2012dynamically,li2014interplay, zhao2021assessing}. This is critical for dynamic dose calculations and motion mitigation strategies~\cite{rana2021investigating}. 
Moreover, accurate modeling of BDT is critical for advanced proton therapy techniques such as PAT, where beam delivery must be synchronized with gantry motion. Robust BDT predictions support efficient treatment planning and trajectory optimization~\cite{mein2024particle}, and are also valuable for investigating dose rate–related toxicities, even in the absence of machine log data~\cite{meijers2024possible}.

To date, the available BDT models have been published only for Ion Beam Applications (IBA) ProteusPLUS~\cite{zhao2022building} and ProteusONE~\cite{zhao2022developing} machines, and the Hitachi particle therapy system~\cite{liang2022investigation, burguete2025stochastic}. No machine-specific BDT model has been reported for Mevion systems.

Machine Learning (ML) has been successfully applied in particle therapy across various domains \cite{wildman2025recent}, including optimizing detector efficiency \cite{missiaggia2022exploratory}, estimating biological damage \cite{cordoni2023artificial}, simulating dose distributions \cite{neishabouri2025real}, and predicting clinical outcomes \cite{padannayil2023impt}. However, its application to beam time delivery prediction remains limited. In this work, we developed the first ML-based Beam Delivery Time (BDT) model for the Mevion S250i machine with Hyperscan technology. The model was trained on machine log files from patients' treatments at our institution. The BDT model was applied to two clinically relevant applications: prediction of BDT for volumetric repainting plans and 4D dose calculation for interplay evaluation. This work aims to fill the gap in the available BDT models and provide insights into the delivery time characteristics of the Mevion machine.

\section{Materials and methods}
\label{MM}

\subsection{Mevion S250i machine and log files}
The Mevion S250i system, equipped with Hyperscan PBS technology, accelerates protons to a fixed energy of around 230~MeV and directs the beam toward the treatment room. To achieve clinically relevant energies, corresponding to penetration depths ranging from 0~cm to 32.2~cm in water, the beam is degraded using the Range Modulator System (RMS), which is mounted on the nozzle. The RMS is composed of 18 Lexan plates of varying thicknesses. By combining these plates, the system can generate 161 distinct energy levels, each separated by 2.1~mm in water-equivalent thickness.

At the distal end of the beamline, mounted on the extendable nozzle, is a dynamic field collimation system known as the Adaptive Aperture (AA). This system reduces lateral penumbra, particularly at lower energies, by trimming the beam laterally within a 20~$\times$~20~cm\textsuperscript{2} area at the isocenter plane. Further details regarding the Mevion S250i system and its commissioning at our institution are available in Vilches et al.~\cite{vilches2020beam}.

For clarity, we define two key terms: \textit{spot} and \textit{pulse}. A \textit{spot} refers to a unique combination of transverse beam position (X, Y) and energy. In the Mevion synchrocyclotron, protons are extracted in discrete \textit{pulses}, short bursts inherent to the extraction process. Each pulse carries a maximum charge of about 8 pC. When the prescribed charge for a spot exceeds this limit, the spot is subdivided into multiple \textit{pulses}. 

After each treatment session, the Mevion system automatically stores machine log files (also referred to as treatment records or dosimetry records), which are currently used for patient-specific quality assurance (PSQA) in our institution. These logs contain more than 300 machine parameters for each delivered pulse, including timestamp, actual and target pulse positions, charge data, and individual AA leaf positions. To deliver a treatment plan, it must first be exported from the treatment planning system (TPS), RayStation, to ARIA, the oncology information system. During this process, Mevion’s proprietary Spot Map Converter (SMC) algorithm reorders the spots, optimizing the machine’s delivery sequence. The resulting plan exported to ARIA mirrors the structure of a machine log file, containing the ordered pulses, target spot positions, AA and RMS configurations, and dose values.

\subsection{The dataset and data process}
The training, cross-validation, and testing of the model are based on machine log files from patients treated in 2025 in Maastro. We randomly selected a maximum of two patients for each indication, restricting the dataset to 11 patients, 1120 treatment records, and 2827772 pulses. Machine log files from different treatment fractions were included in the dataset to account for inter-fractional delivery uncertainties.

To ensure robust model performance and meaningful feature representation, a comprehensive data preparation pipeline was implemented. Since the model predicts the beam delivery time before irradiation, we selected features that are available before treatment delivery: pulse target charge, pulse position at the isocenter, AA position, and beam energy.
We then computed the difference, $\Delta \Theta_{n} = \Theta_{n}-\Theta_{n-1}$, between the n-th and the previous pulse ($n-1$) for each feature $\Theta$. In particular, we summarized the AA information by calculating the sum of all the AA leaves shifted as follows:
$$\Delta AA = \sqrt{\sum_{l=1}^{14}\Delta X^2_{l}+\Delta Y^2_{l}}$$
where $l$ is the $l$-th AA leaf and $X$ and $Y$ are the coordinates of the leaves at the isocenter. This formulation assumes that the $\Delta T$ depends on the magnitude of the AA shift to move from one spot to the other one. The same principle was applied to the spot position $\Delta S$, where the Cartesian length of two consecutive pulses was used as model features instead of the single coordinates.
Two additional parameters were derived from the data: \textit{isFirstPulse} and \textit{isTxPulse}. 
The former indicates the first pulse delivered for a given beam and treatment fraction, whereas the latter marks the first pulse following the so-called `low charge' layer. At the beginning of each beam irradiation, the machine delivers the first energy layer defined in the sequence, setting 5~pC for each pulse. After that, the machine checks the spot positions and applies a systematic shift correction on all the remaining energy layers to correct the spot positions. This process could introduce time delays between the last low charge pulse and the first treatment pulse. Starting from these features, we derived the following additional variables.

\textit{Log-Transformed Features:} To mitigate skewness in the distributions of key continuous variables, logarithmic transformations were applied:
\begin{itemize}
    \item \texttt{$log_{\Delta AA}$ } = $\log(1 + \texttt{$\Delta AA$})$
    \item \texttt{$log_{\Delta S}$} = $\log(1 + \texttt{$\Delta S$})$
    \item \texttt{$log_{\Delta E}$} = $\log(1 + |\texttt{$\Delta E$}|)$
\end{itemize}

\textit{Interaction Features:} Interaction terms were introduced to capture compound effects:
\begin{itemize}
    \item \texttt{AA\_Spot\_Interaction} = \texttt{$\Delta AA$} $\times$ \texttt{$\Delta S$}
    \item \texttt{Energy\_AA\_Interaction} = $|\texttt{$\Delta E$}| \times \texttt{$\Delta AA$}$
    \item \texttt{Energy\_Spot\_Interaction} = $|\texttt{$\Delta E$}| \times \texttt{$\Delta S$}$
\end{itemize}

\textit{Categorical Binning} To model non-linear relationships, continuous variables were discretized into categorical bins:
\begin{itemize}
    \item \texttt{Energy\_Category}: Binned into \texttt{Zero} (0-0.1 MeV), \texttt{Low} (0.1-50 MeV), \texttt{Medium} (50-100 MeV), \texttt{High} (100-150 MeV), and \texttt{VeryHigh} ($>$150 MeV).
    \item \texttt{AA\_Category}: Binned into \texttt{Zero} (0-0.1 mm), \texttt{Small} (0.1-10) mm, \texttt{Medium} (10-100 mm), and \texttt{Large} ($>$ 100 mm).
\end{itemize}

\textit{Boolean Indicators:} Binary features were created to flag specific conditions:
\begin{itemize}
    \item \texttt{Is\_Energy\_Change}: Indicates energy variation ($> 0$ MeV)
    \item \texttt{Is\_Major\_AA\_Change}: Indicates substantial aperture adjustment ($> 2$ mm)
    \item \texttt{Is\_Major\_Spot\_Change}: Indicates notable spot position change ($> 1$ mm)
\end{itemize}

\textit{Movement Metrics:} A composite movement indicator was defined to quantify overall physical change:
\begin{align*}
    \texttt{Total\_Movement} &= \sqrt{\texttt{$\Delta AA$}^2 + \texttt{$\Delta S$}^2 + |\texttt{$\Delta E$}|^2} \\
    \texttt{log\_Total\_Movement} &= \log(1 + \texttt{Total\_Movement})
\end{align*}

The target variable $y$ is defined as the inter-pulse time delta, \texttt{$\Delta T$}. It was log-transformed to reduce skewness and improve model sensitivity:$$ y_{\text{log}} = \log(1 + y)$$.

A stratified train-test split was performed using quantile-based binning of the log-transformed target to preserve its distributional characteristics across subsets. The final split allocated 70\% of the data to training and 30\% to testing, ensuring balanced representation of temporal dynamics.

The list of features used as input for the ML model is provided in the supplementary materials in table~\ref{tab:features}. 

\subsection{Random Forest: training, cross-validation and testing}
We employed a Random Forest (RF) to model the relationship between the inter-pulse time difference $\Delta T$ and the input features. Random Forest is an ensemble learning method that constructs multiple decision trees and aggregates their predictions in parallel to improve accuracy and reduce overfitting. This approach was chosen for its robustness to noise and ability to handle nonlinear relationships~\cite{ho1995random, ho1998random}. The model was implemented in Python (v3.12.3) using the machine learning library scikit-learn (v1.6.1)\cite{scikit-learn}. The pipeline consists of two main components: a preprocessing stage and a regression model. 

The complete preprocessing pipeline included:
\begin{itemize}
    \item \textit{Robust Scaling:} This Scaler removes the median and scales the numeric variables according to the quantile range: between the 1st quartile (25th quantile) and the 3rd quartile (75th quantile).
    \item \textit{Passthrough:} Binary categorical features were retained in their original form.
    \item \textit{Ordinal Encoding:} Applied to discretized categorical features using an ordinal encoder with unknown value handling.
\end{itemize}

The transformed features were used to train a RF. To further refine the model, a randomized hyperparameter search was conducted using 5-fold cross-validation. 

\subsection{Error estimation}
To evaluate the performance of the optimized RF, we utilized relevant error metrics specific to our application. The Mean Absolute Error (MAE) $$MAE=\frac{1}{N}\sum_{i=0}^N |\Delta T^*_i - \Delta T_i|$$
quantifies the average magnitude of prediction errors, providing an intuitive measure of accuracy. Additionally, the Mean Absolute Percentage Error (MAPE) metric was used to estimate the relative error to the predicted $\Delta T$ with respect to the reference values, $\Delta T^*$:
$$MAPE=\frac{1}{N}\sum_{i=0}^N \frac{|\Delta T^*_i - \Delta T_i|}{\Delta T^*_i}$$. We also divided the range of $\Delta T$ into sub-ranges (0-50 ms, 50-500 ms, 500-1000 ms, $>$1000 ms), and we evaluated the model performances on each interval.

\subsection{Explainable AI}

To interpret the predictions of the machine learning (ML) model, we employed the SHAP (SHapley Additive exPlanations) method. SHAP is a post-hoc explainable artificial intelligence (XAI) approach based on Shapley values from cooperative game theory~\cite{lundberg2017unified}. It attributes the contribution of each input feature to a specific model output by quantifying how the prediction changes when the feature is included versus excluded in different coalitions of features.

Formally, SHAP computes Shapley values as an additive feature attribution model, analogous to a linear model, such that the prediction is decomposed as
\[
\hat{f}(x) = \phi_0 + \sum_{i=1}^n \phi_i\,,
\]
where $\phi_i$ represents the contribution of the $i$-th feature, and $\phi_0$ is the expected value of the model output over the training data (i.e., the base value). The SHAP value $\phi_i$ for feature $i$ is defined as:
\[
\phi_i = \sum_{S \subseteq N \setminus \{i\}} \frac{|S|!(|N|-|S|-1)!}{|N|!} \left[ f(S \cup \{i\}) - f(S) \right]
\]
where $N$ is the set of all input features, $S$ is a subset of features not containing $i$, and $f(S)$ is the model prediction using only the features in $S$.

We specifically employed TreeSHAP, a tree-based SHAP algorithm designed for efficient and accurate SHAP value computation in tree-based models such as Random Forest (RF)~\cite{lundberg2020local}. TreeSHAP leverages the structure of decision trees to compute exact SHAP values in polynomial time, ensuring consistent and locally accurate explanations of individual predictions.

In this study, SHAP values were computed for 100 data samples from the test set to cover the $\Delta T$ range. Global explanations were obtained by analyzing individual feature contributions and by grouping features with similar physical meaning across the dataset, highlighting the most influential features for model performance. The feature groups used in the SHAP analysis are listed below.

The SHAP implementation was performed using the open-source Python library \texttt{shap} (v0.48.0), applied to the model trained with optimal hyperparameters selected using the procedure described earlier.

\subsection{Clinical applications: volumetric repainting and interplay evaluation}
We applied the proposed time prediction model to two clinically relevant scenarios where delivery time estimation is valuable in the absence of log file data: (i) 4D dose calculation for interplay evaluation and (ii) dose rate estimation.

One of the most clinically significant applications of the BDT model is the evaluation of interplay. To assess model accuracy, we predicted the beam delivery time (BDT) for a lung cancer treatment plan optimized with five volumetric repaintings. Since this configuration is not routinely used in clinical practice, a dry run was performed to generate machine log files for reference.

To quantify the dosimetric impact of model-based time predictions on 4D dose calculations, we simulated the interplay effect by combining the ML-based BDT model with 4DCT data and a synthetic breathing trace. This approach allowed mapping each treatment pulse to the corresponding 4DCT phase \cite{arxivInterplay}. For this analysis, we used the clinical plan delivered without volumetric repainting. The same procedure was repeated using machine log files from the first ten treatment fractions to verify whether the model uncertainty remains within the intrinsic variability of machine delivery. In clinical practice, volumetric repainting is not used; therefore, this plan was created without incorporating any target repainting techniques.

\section{Results}
\label{Res}

\subsection{Inter-Pulse Timing Exploratory Analysis}

\begin{figure}
    \centering
    \includegraphics[width=1\textwidth]{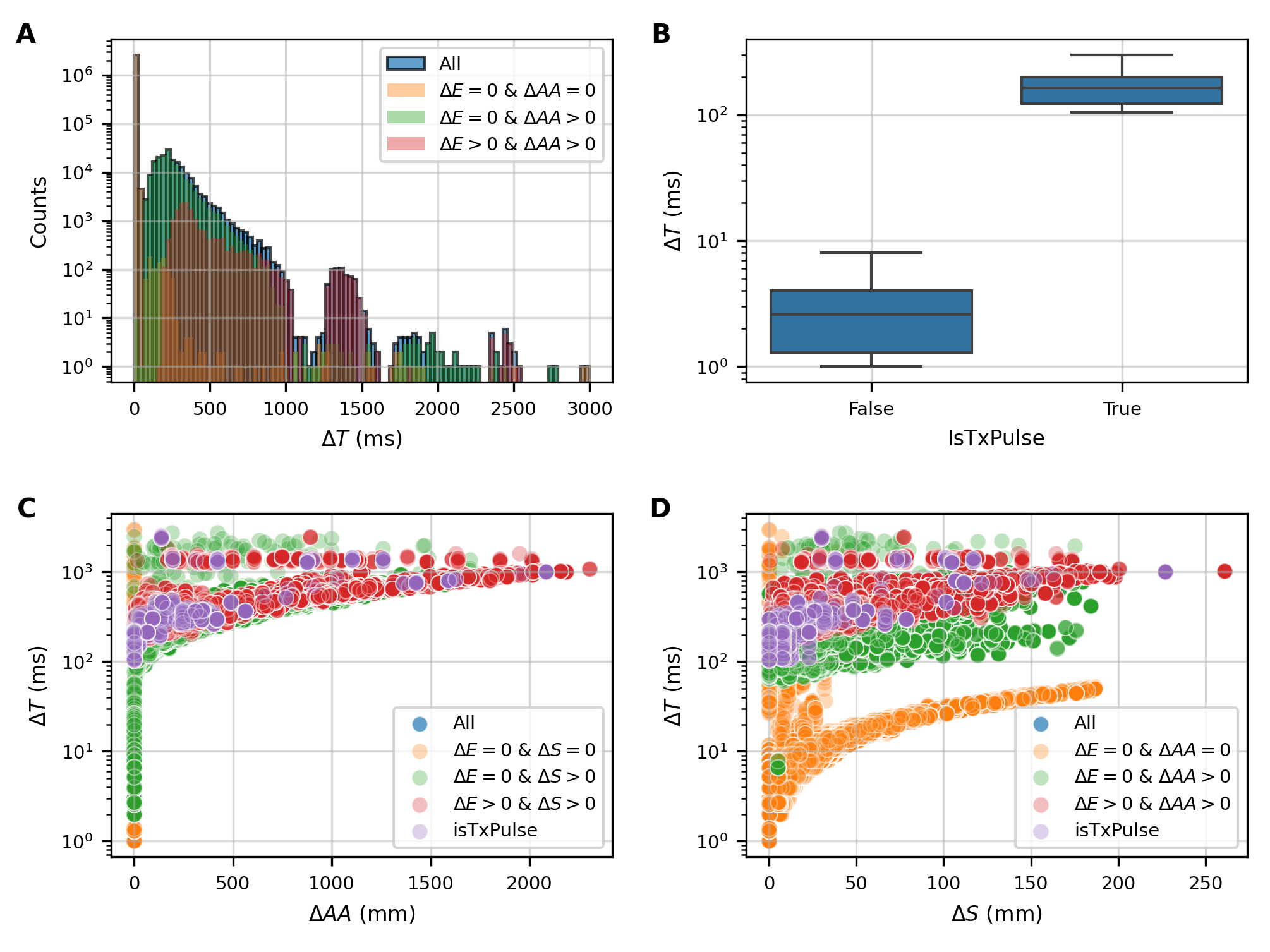} 
    \caption{Impact of energy and spatial changes on delivery time. (A) Event counts across $\Delta T$ intervals categorized by energy and spatial changes. (B) Box plot of $\Delta T$ distribution based on the \texttt{IsTxPulse} condition. (C) Scatter plots showing relationships between AA change, and $\Delta T$ under different conditions, including cases where energy or spot position changes are zero or positive/negative, with \texttt{IsTxPulse} highlighted. (D) same as panel (C) with $\Delta T$ -$\Delta S$ dependence.}
    \label{fig:TimeStamp}
\end{figure}

Figure~\ref{fig:TimeStamp} illustrates the heterogeneity of the inter-pulse time intervals ($\Delta T$) observed in the Mevion S250i system. Figure~\ref{fig:TimeStamp}(A) presents a histogram of $\Delta T$ values across the dataset, showing a broad distribution ranging from a few milliseconds up to approximately $10^3$~ms. We grouped $\Delta T$ according to relevant machine features: the orange bars represent the inter-pulse time distribution when no energy switching ($\Delta E=0$ MeV) and no AA shifts ($\Delta AA=0$ mm) occur, the green bar for pulses when AA is moving within the same energy layer ($\Delta AA>0$ mm and $\Delta E=0$ MeV), and the red bars when range shifter plates and AA are moving. The shortest inter-pulse intervals occur when no energy switching or AA adjustment is required, specifically, when the beam delivers pulses within the same energy layer.

Figure~\ref{fig:TimeStamp}(B) shows the box plot of $\Delta T$ values grouped by \textit{IsTxPulse} variable. Pulses following the low-charge layers (\textit{IsTxPulse=True}) exhibit systematically longer $\Delta T$ values compared to subsequent pulses. 

Figure~\ref{fig:TimeStamp}(C) shows the relation between $\Delta T$ and the magnitude of the spot shift ($\Delta AA$). Similarly, the panel (D) highlights the relation between spot position shifts ($\Delta \mathrm{S}$) and $\Delta T$. The movement of machine components, such as AA shifts, introduces a significant increase in inter-pulse dead time of around an order of magnitude.

\begin{figure}
    \centering
    \includegraphics[width=1\textwidth]{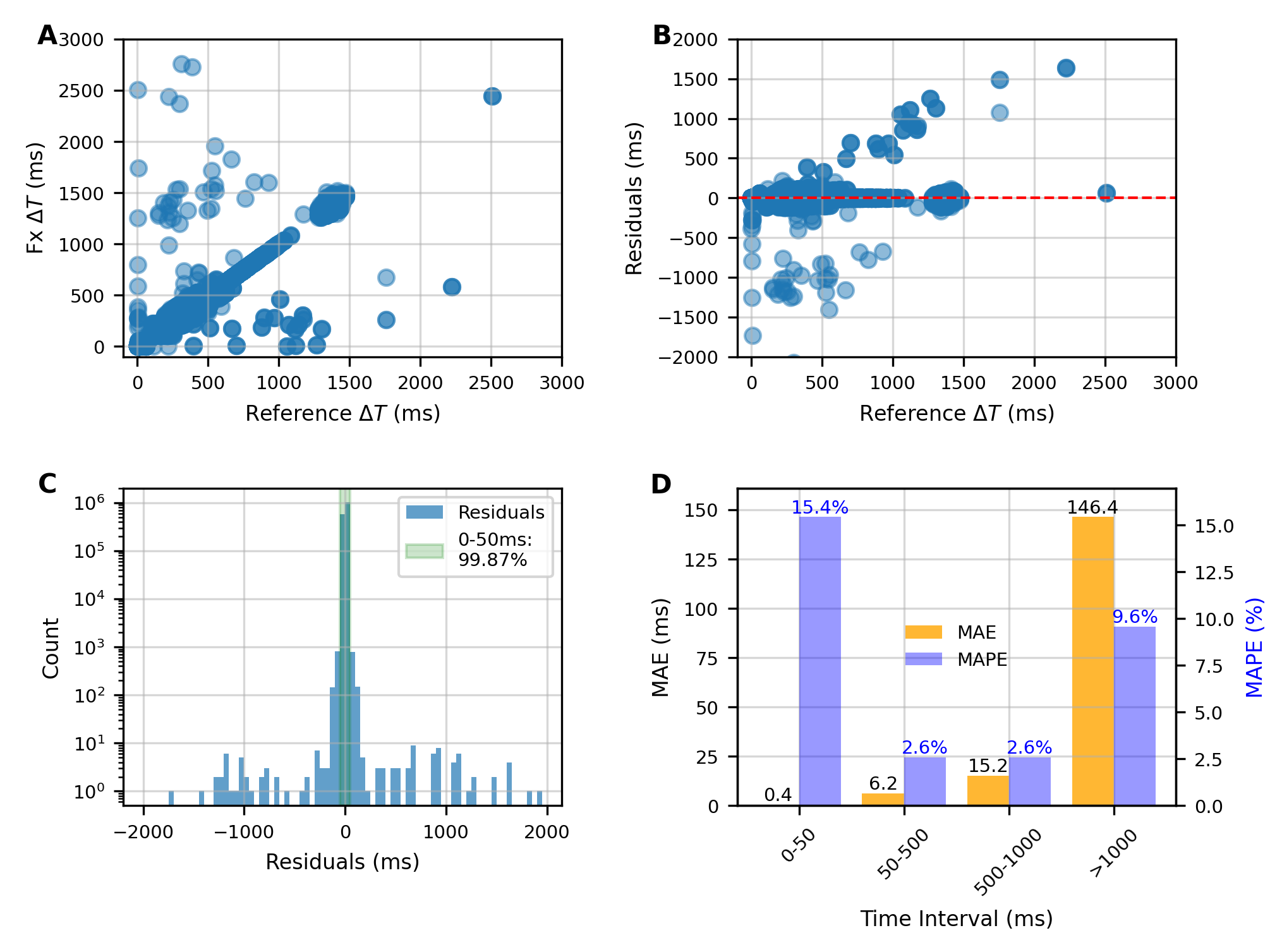}
   \caption{Analysis of delivery time variability across treatment fractions. (A) Fractional $\Delta T$ (ms) versus the first fraction $\Delta T$ (ms), as reference. (B) Residuals (ms) versus reference $\Delta T$ (ms). (C) Histogram of residual counts across intervals, highlighting the range $\pm  50$ ms. (D) Mean absolute error (MAE) in ms across different time intervals.}
    \label{fig:TRVariability}
\end{figure}

Figure~\ref{fig:TRVariability} shows the intrinsic log file variability in delivering the same plan over the treatment fractions. The $\Delta T$ values from the first fraction were used as a reference, and we compared them with the delivery times recorded on the next fractions. Panel (A) shows a scatter plot comparing the fractional $\Delta T$ against the reference $\Delta T$. A small dataset largely deviates from the diagonal. Figure~\ref{fig:TimeStamp}(B) displays the residuals (difference between reference and single fraction $\Delta T$) plotted against the reference values. The red dashed line at zero highlights the ideal case of no error. The largest amount of data is spread within 100-200 ms. The machine shows a systematic reproducibility in delivering the spots between 500 and 1000 ms, a time interval that is systematically aligned to the diagonal with the smallest residual values.
Panel (C) presents a histogram of residuals, with the y-axis on a logarithmic scale to emphasize the distribution of error magnitudes. A green line marks the ±500~ms band, within which 99.87\% of the predictions fall. To understand the log file delivery uncertainty, we computed the MAE and the MAPE across different $\Delta T$ intervals (Figure~\ref{fig:TimeStamp}(D)). The absolute error increases with longer time intervals, ranging from 0.4~ms in the 0–50~ms range to 146.4~ms for intervals exceeding 1000~ms. Conversely, the MAPE shows the highest value at low $\Delta T$, 15.4 \% within 50 ms, while for $\Delta T>1000$ ms the percentage error is around 10\%.

\subsection{Model performance}
\begin{figure}
    \centering
    \includegraphics[width=\linewidth]{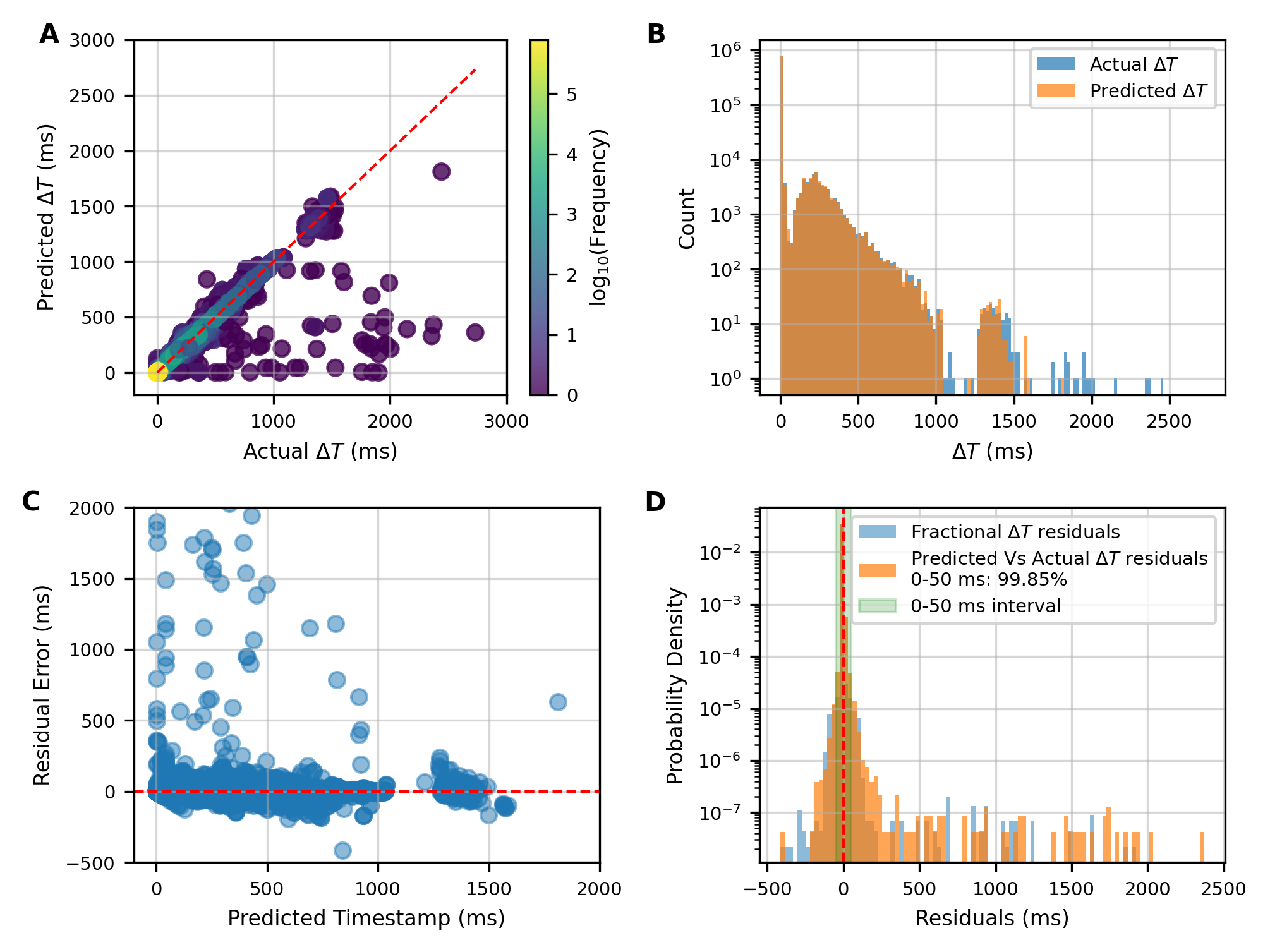}
   \caption{Evaluation of model performance and residual distribution. (A) Predicted $\Delta T$ (ms) versus actual $\Delta T$ (ms), colored by event density. (B) Histogram of event counts across $\Delta T$ intervals for actual and predicted values. (C) Residual error (ms) versus predicted timestamp (ms). (D) Probability density of residuals with highlighted 0-50 ms interval.}
    \label{fig:ModelResulstTest}
\end{figure}

Figure~\ref{fig:ModelResulstTest} presents a detailed evaluation of the model's ability to predict inter-pulse time intervals ($\Delta T$) on the test dataset.

Figure~\ref{fig:ModelResulstTest}(A) shows a scatter plot of predicted versus actual $\Delta T$ values. Most data points align closely with the red dashed identity line, indicating strong agreement between predictions and ground truth. The color gradient highlights the density of the data. A small number of points below the diagonal suggests an underestimation of the predicted $\Delta T$ values in specific cases. Panel (B) compares the distributions of actual and predicted $\Delta T$ values using overlaid histograms. The orange bars represent the predicted distribution, while the blue bars correspond to the reference values. The model successfully captures the three characteristic regions: a sharp peak below 50~ms (no AA or energy change), a mid-range from 50–1000~ms (primarily AA movement), and a long tail beyond 1000~ms (combined AA and RMS shifts).

Panel (C) represents the residuals as a function of predicted timestamps. Most residuals cluster around zero, with the red dashed line indicating the ideal case of no error.
Finally, the bottom right panel (D) shows the probability density function of residuals with orange bars. Blue bars represent the intrinsic machine log file residuals due to inter-fractional delivery uncertainty, already shown in Figure~\ref{fig:TRVariability}, as a reference. The orange histogram highlights that the majority of errors fall within a narrow band around zero, in agreement with the intrinsic machine delivery uncertainties.
Below are reported the MAE and MAPE to quantify the model accuracy per $\Delta T$ interval:
\begin{itemize}
    \item \textit{0-50 ms:} MAE=0.9 ms / MAPE=36\%
    \item  \textit{59-500 ms:} MAE=6.1 ms / MAPE=2.5\% 
    \item \textit{500-1000 ms:} MAE=10.7 ms / MAPE=1.7\% 
    \item \textit{$\ge$1000 ms:} MAE=222.0 ms / MAPE=13.2\%
\end{itemize}
These values align with the intrinsic machine BDT uncertainty (Figure~\ref{fig:TRVariability}(D)), except for the low time interval where the model indicates double MAE and MAPE values.

\subsection{Explainable AI}
\begin{figure}
    \centering
    \includegraphics[width=1\linewidth]{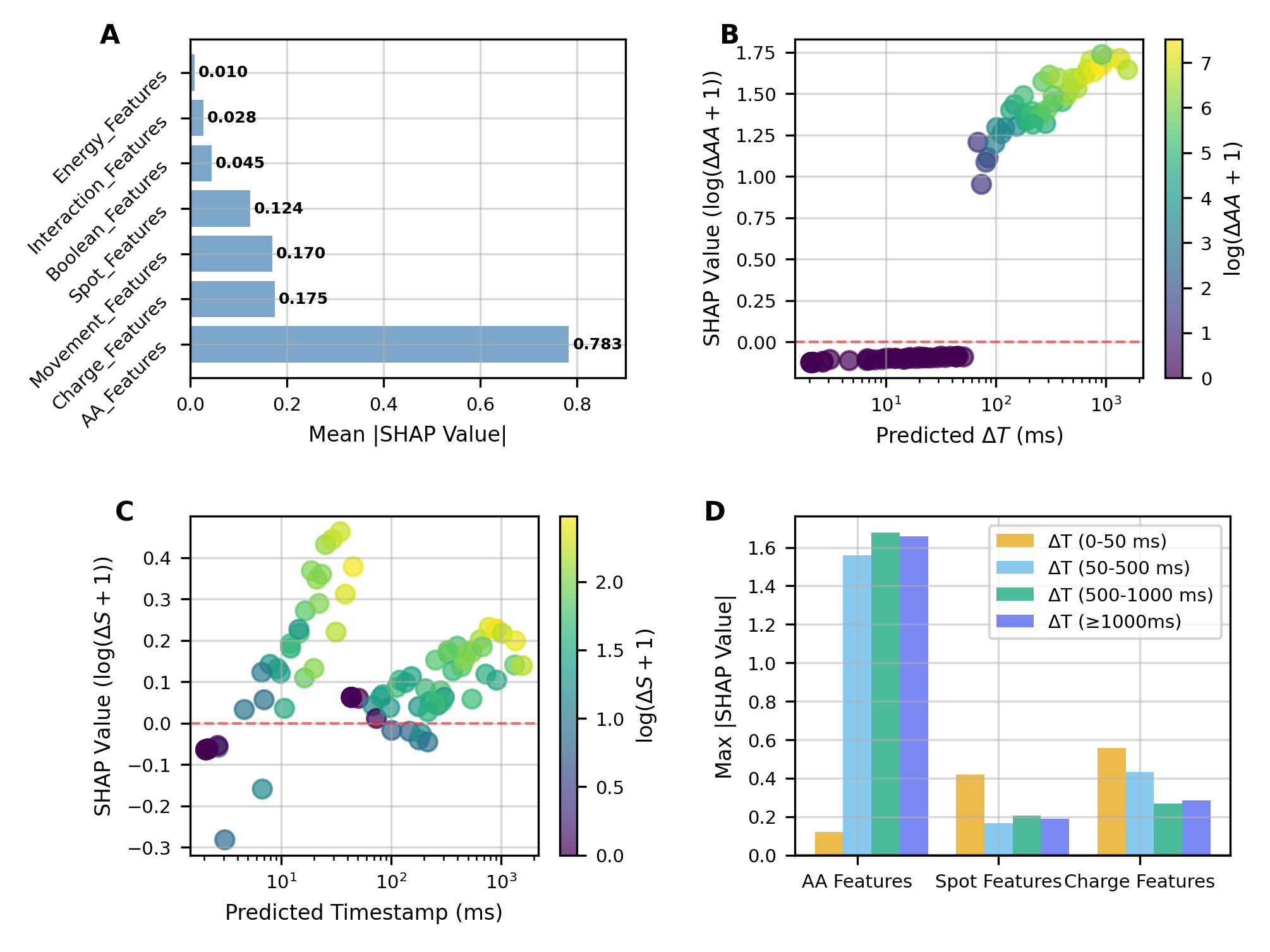}
    \caption{Comprehensive SHAP analysis of the BDT model. (A) Mean absolute SHAP values for different feature groups: energy, interaction, boolean, spot, movement, charge, and AA features. (B) SHAP value for $\log(\Delta AA)$ versus predicted $\Delta T$ (ms), colored by $\log(\Delta AA + 1)$. (C) SHAP value for $\log(\Delta S)$ versus predicted timestamp (ms), colored by $\log(\Delta S + 1)$. (D) Maximum SHAP values for AA, spot, and charge features across different $\Delta T$ intervals.}
    \label{fig:shap}
\end{figure}

We computed SHAP values to estimate both global and local feature contributions to the model's predictions. Figure~\ref{fig:shap} summarizes the SHAP analysis. Panel (A) shows that AA features are globally the most influential, with a mean absolute value SHAP value of 0.78, followed by Charge (0.175), Movement (0.170), and Spot (0.124) features. The scatter plot in panel (B) reveals a strong local correlation between $\log_{\Delta \mathrm{AA}}$ and predicted $\Delta T$, with SHAP values sharply increasing above 70 ms from -0.17 to a maximum of 1.75. Panel (C) shows that $\log_{\Delta S}$ contributes variably across predictions, indicating its relevance when AA shifts are low. Figure~\ref{fig:shap}(D) compares feature importance across $\Delta T$ intervals. AA features dominate across all intervals above 50 ms, with a maximum SHAP value ranging from 1.56 to 1.68. Spot and Charge features show higher importance than AA features for $\Delta T$ < 50 ms.  Movement features were excluded due to their direct dependence on Spot and AA shifts.

\subsection{Clinical Applications: Repainting and Interplay evaluation}
A relevant clinical application of the BDT model is predicting the single-pulse delivery time during interplay evaluation. We first validated the model by reproducing the treatment time of a lung cancer plan optimized for delivery with five volumetric repaintings. In this approach, the target is irradiated five times within a treatment fraction to mitigate the interplay effect.

Figure~\ref{fig:TimeStamp}(A) compares the predicted $\Delta T$ values with those extracted from the machine log file. The data points align closely along the diagonal, representing the ideal model trend, which indicates good predictive accuracy. However, predicted $\Delta T$ values exceeding $1000~\mathrm{ms}$ are systematically underestimated compared to the actual measurements. 

This systematic deviation is further illustrated in panel (B), where the cumulative $\Delta T$ over the entire treatment is plotted against the prescribed energy. The plot highlights the temporal energy sequence characteristic of this plan: during the first repaint, all energies are delivered, followed by four repetitions from the lowest to the highest energy. The model successfully reproduces the characteristic dead times associated with energy plate adjustments and AA movements, both within a repaint and when transitioning to the next. The cumulative $\Delta T$ plot reveals that the systematic underestimation at high $\Delta T$ values propagates throughout the treatment, resulting in a consistent shift in cumulative delivery time relative to the actual irradiation time. We estimate the final discrepancy to be approximately $-1.7\%$ (predicted minus actual), corresponding to a difference of about $4~\mathrm{s}$ over a $5$ minutes irradiation per beam.
\begin{figure}
    \centering
    \includegraphics[width=1\linewidth]{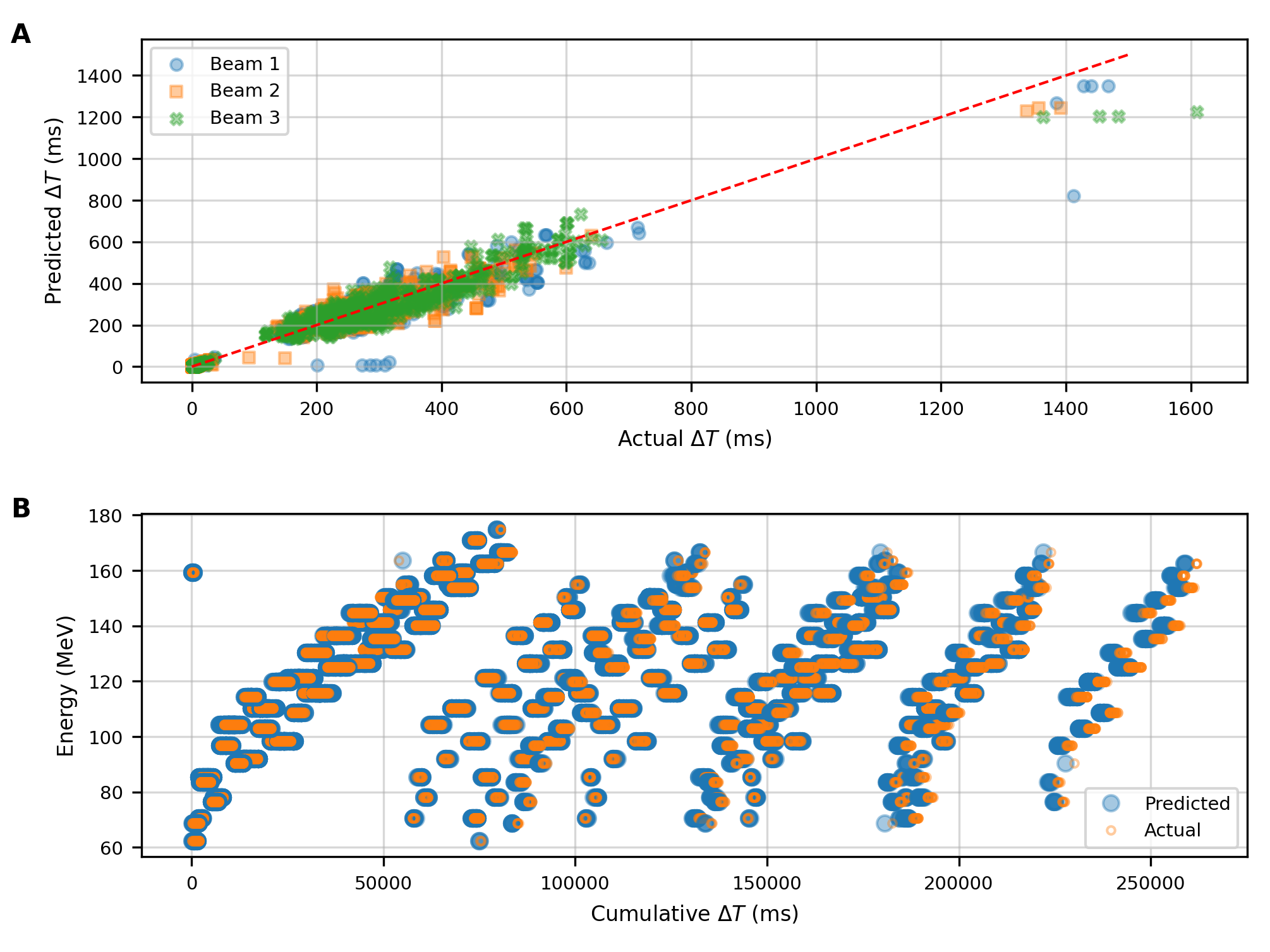}
   \caption{Comparison between predicted and actual delivery times for a lung cancer treatment plan with five volumetric repaintings. (A) Predicted $\Delta T$ (ms) versus actual $\Delta T$ (ms) for three beams. The diagonal line represents the ideal model trend. (B) Cumulative $\Delta T$ as a function of energy (MeV) for predicted and actual $\Delta T$ values.}
    \label{fig:repainting}
\end{figure}

\begin{figure}
    \centering
    \includegraphics[width=1\linewidth]{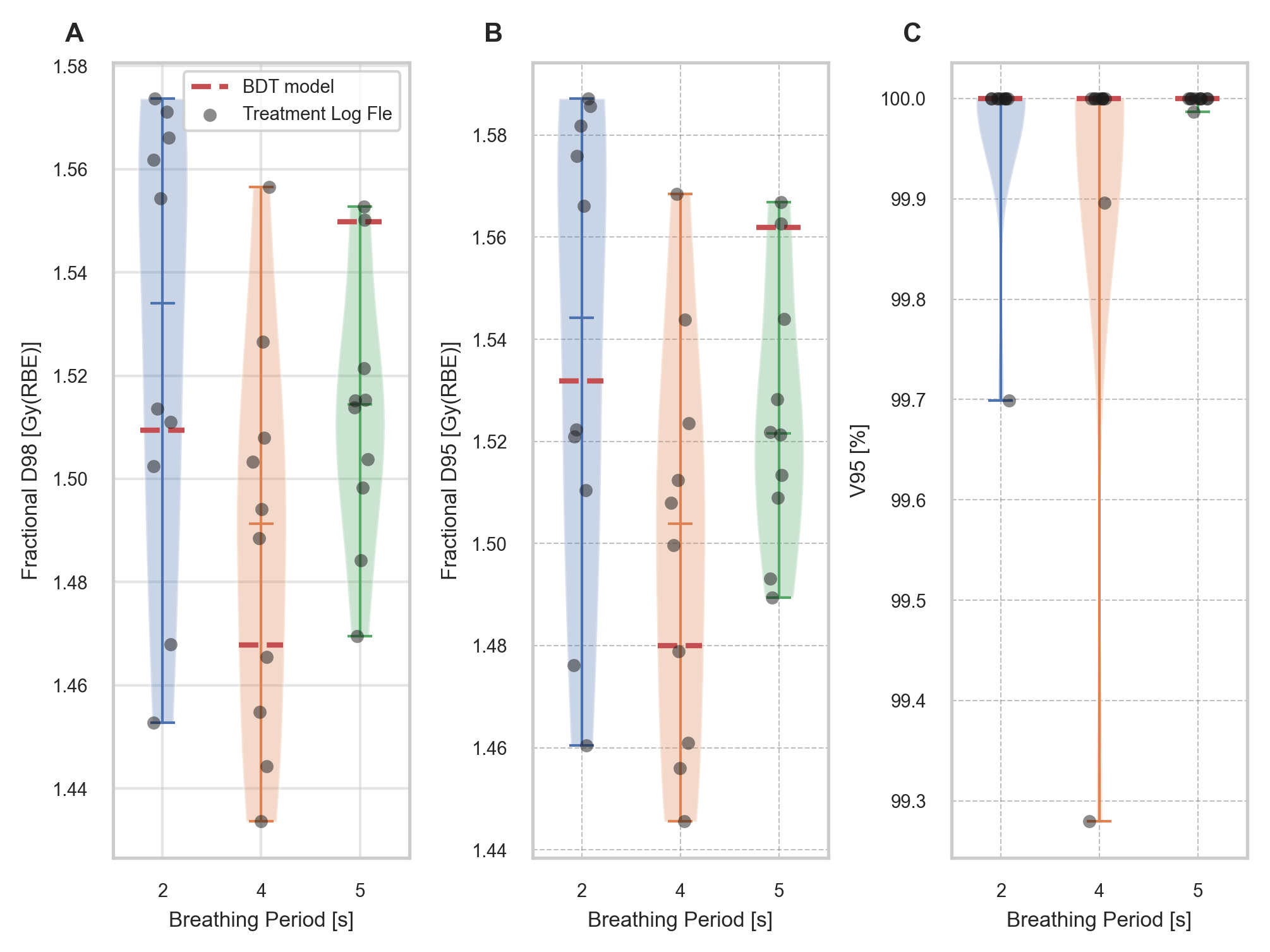}
    \caption{Comparison of dosimetric metrics from a 4D dose calculation using the BDT model and actual machine log files for interplay evaluation. Panels (A), (B), and (C) show the violin plots of the D$_{98}$, D$_{95}$, and V$_{95}$, respectively, across different breathing periods (2~s, 4~s, and 5~s). Black dots represent values from treatment log files of the first ten fractions, while the red dashed line indicates BDT model predictions.}
    \label{fig:interplay}
\end{figure}

Figure~\ref{fig:interplay} illustrates the dosimetric impact of using model-predicted delivery times to recalculate the 4D dose distribution for interplay evaluation. The 4D dose was computed by combining the machine time model with a sinusoidal breathing trace of periods $2~\mathrm{s}$, $4~\mathrm{s}$, and $5~\mathrm{s}$. For comparison, the same calculation was performed using actual machine log files from the first ten treatment fractions. Since the actual patient-delivered plans were used in this analysis, no volumetric repainting was applied.

Panels~A, B, and C show that the D$_{98}$, D$_{95}$, and V$_{95}$ values obtained with the BDT model predictions remain within the intrinsic delivery variability of the machine for all breathing periods considered. 

\section{Discussion}
\label{Disc}

In this study, we developed and validated the first ML-based BDT prediction model for the Mevion S250i proton therapy system with Hyperscan technology. Using institutional machine log files, we trained a Random Forest model to predict inter-pulse delivery times ($\Delta T$). Our results show that the model successfully reproduces the characteristic temporal dynamics of the Mevion system, capturing both the wide range and the complex distribution of inter-pulse times. 

Our exploratory analysis (Figure~\ref{fig:TimeStamp}) highlighted three key determinants of inter-pulse time: (i) energy layer switching, (ii) the magnitude of AA and spot position movements, and (iii) transitions from low-charge layers to treatment pulses. Compared to prior BDT studies on IBA and Hitachi systems \cite{zhao2022building, zhao2022developing}, this analysis underscores Mevion-specific characteristics, particularly the RMS-based energy selection and AA collimation. For example, while in the IBA ProteusONE, the spot-switching time dominates delivery time \cite{zhao2022developing}, in the Mevion S250i, the combined effects of AA movement and RMS switching represent the main bottleneck.

This is also the first study to employ a machine-learning approach to model the BDT of a proton machine. Our RF model achieved predictive accuracies across all $\Delta T$ intervals comparable with the daily machine delivery uncertainties (Figure~\ref{fig:TRVariability}). The largest percentage errors were observed for very short inter-pulse times ($<50$~ms), around 35\%, which is approximately double the intrinsic daily variability. However, relative errors are not an appropriate metric in this range, as dividing by small numbers inflates the percentage deviation. In fact, the MAE was only 0.9 ms, a negligible uncertainty compared with the overall range of $\Delta T$ values.

\subsection{Explainable AI}
The SHAP analysis (Figure~\ref{fig:shap}) revealed that AA-related features were globally the most influential predictors. However, AAs were not relevant across all $\Delta T$ values: for short inter-pulse times ($<50$~ms), charge and spot position features dominated. This is consistent with the exploratory analysis in Figure~\ref{fig:TimeStamp}, which showed that $\Delta T$ values below 50--70 ms were primarily determined by spot switching within the same energy layer and AA configuration. For longer inter-pulse times, $\Delta T$ increased due to AA leaf shifts, reflected in higher SHAP values for AA-related features. 

In contrast to the exploratory analysis, the boolean variable \textit{IsTxPulse} contributed only marginally to model predictions. This likely reflects the fact that, after completing a low-charge layer, the system typically advances to a new AA or energy layer, introducing systematic delays already captured by other features. Similarly, energy-related features such as energy deltas and \textit{IsEnergyChange} had a small global contribution. At the local level, however, SHAP analysis indicated that energy switching ($\log_{\Delta E}$) gained importance at higher $\Delta T$ values, scaling with the magnitude of $\Delta E$ (see Figure~\ref{fig:shapEnergy} in the supplementary materials). This is physically plausible given the RMS design: larger energy steps require repositioning more range shifter plates, introducing additional dead time. Still, the maximum absolute SHAP value for this feature was only 0.26, which is lower than for the AA, spot, and charge features, indicating that the model primarily uses energy-related features for prediction fine-tuning rather than as dominant predictors.

\subsection{Clinical Applications}
The BDT model was applied to two clinically relevant scenarios. First, in volumetric repainting of a lung plan, the model reproduced the overall timing pattern even when the machine restarted a new repaint. Since training data did not include volumetric repainting, this result suggests good generalization to more complex delivery structures. The model slightly underestimated long delays, resulting in a cumulative deviation of $-1.7\%$. The selected case, which involved an unusually high number of repaintings, represents a worst-case scenario not typically encountered in clinical practice. This highlights that error propagation across extended delivery times remains small and unlikely to be clinically significant. 

Second, model-predicted delivery times were integrated into 4D dose calculations. The resulting D$_{98}$, D$_{95}$, and V$_{95}$ values closely matched those derived from machine log files, showing that model uncertainty is within the intrinsic variability of the Mevion system.

\subsection{Limitations}
This study highlights the key delivery characteristics of the Mevion S250i; however, the dataset is limited to a single institution. Multi-institutional studies are necessary to assess the generalizability and reproducibility of these findings across the broader Mevion user community. The model also underestimated rare long-delay events, suggesting more sophisticated data processing, such as oversampling strategies, could improve predictive accuracy. While Random Forests were selected for their robustness and interpretability, alternative approaches such as gradient boosting \cite{friedman2001greedy} or neural networks may further enhance performance.

Future work should extend this framework to emerging modalities such as PAT, where precise synchronization of BDT is critical for plan optimization \cite{wuyckens2025proton}. More broadly, incorporating BDT prediction into treatment planning systems could enable delivery time–based optimization for a variety of applications, including organ motion management and overall delivery time reduction \cite{butkus2025spot}. With the growing adoption of hypofractionated treatments, machine-specific modeling of delivery time may also help mitigate interplay effects \cite{santos2022role}. Finally, accurate BDT prediction supports throughput estimation, which is increasingly important for patient scheduling and resource optimization in high-volume proton therapy centers \cite{liu2016methods}.

\section{Conclusions}

We developed and validated the first machine learning–based beam delivery time (BDT) model for the Mevion S250i Hyperscan system. The Random Forest model accurately predicts inter-pulse delivery times, capturing the characteristic temporal dynamics of the machine. SHAP analysis indicated that AA movements are the primary drivers of BDT, whereas spot positions, pulse charge, and energy changes contribute variably depending on the inter-pulse interval. Beyond predictive performance, the study provides detailed insights into the machine’s operational characteristics, including the influence of energy layer switching and aperture adjustments on delivery time. The model generalized well to clinical scenarios, such as volumetric repainting and 4D dose calculations, showing cumulative delivery time deviations of only -1.7\% and dosimetric metrics within the intrinsic machine variability.

\appendix
\section{Supplementary Materials}
\label{app1}

\subsection{Random forest BDT model}
\noindent 
Table~\ref{tab:features} summarizes the names, type and description of the features used as input for training the RF model. 
\begin{table}[htbp]
\centering
\begin{tabular}{lcp{8cm}}
\hline
\textbf{Feature} & \textbf{Type} & \textbf{Description} \\
\hline
$\Delta AA$ (mm) & n & Adaptive aperture position difference between pulses\\
$\Delta S$ (mm) & n & Spot position difference between pulses \\
$\Delta E$ (MeV) & n & Pulse energy in mega-electronvolts \\
$\Delta c$ (pC) & n & Target pulse charge in picoCoulombs \\
$c$ (pC) & n & Original target pulse charge before adjustment \\
IsFirstPulse & b & Indicates if the pulse is the first in a sequence \\
IsTxPulse & b & Indicates if the pulse is a transmission pulse \\
$log_{\Delta AA}$ & t & Log-transformed aperture difference \\
$log_{\Delta S}$ & t & Log-transformed spot position difference \\
$log_{\Delta E}$ & t & Log-transformed absolute energy value \\
AA\_Spot\_Interaction & i & Product of aperture and spot position differences \\
Energy\_AA\_Interaction & i & Product of energy and aperture difference \\
Energy\_Spot\_Interaction & i & Product of energy and spot position difference \\
Energy\_Category & c & Binned energy levels: Zero to VeryHigh \\
AA\_Category & c & Binned aperture differences: Zero to Large \\
Is\_Energy\_Change & b & Indicates significant energy change ($> 0.1$ MeV) \\
Is\_Major\_AA\_Change & b & Indicates major aperture change ($> 1$ mm) \\
Is\_Major\_Spot\_Change & b & Indicates major spot position change ($> 1$ mm) \\
Total\_Movement & t & Composite movement metric from AA, spot, and energy \\
log\_Total\_Movement & t & Log-transformed total movement \\
\hline
\end{tabular}
\caption{The tables summarize the names, types, and descriptions of the features used to train the RF model. n = Numeric
t = Transformed (e.g., log or composite)
i = Interaction
b = Boolean
c = Categorical}
\label{tab:features}
\end{table}

Table~\ref{tab:rf_params} shows the best values of the RF hyperparameters selected from the five-fold cross-validation process.

\begin{table}[h]
\centering
\begin{tabular}{ll}
\hline
\textbf{Parameter} & \textbf{Value} \\
\hline
\texttt{bootstrap} & \texttt{True} \\
\texttt{max\_depth} & 31 \\
\texttt{max\_features} & \texttt{0.7} \\
\texttt{max\_samples} & 0.99 \\
\texttt{min\_samples\_leaf} & 4 \\
\texttt{min\_samples\_split} & 7 \\
\texttt{n\_estimators} & 291 \\
\hline
\end{tabular}
\caption{Random Forest hyperparameter values after 5-fold cross-validation tuning.}
\label{tab:rf_params}
\end{table}

\subsection{Explainable AI: Energy feature}
\noindent SHAP value of $log_{\Delta E}$ feature as a function of the predicted $\Delta T$. The colormap highlights the feature value.
\begin{figure}
    \centering
    \includegraphics[width=1\linewidth]{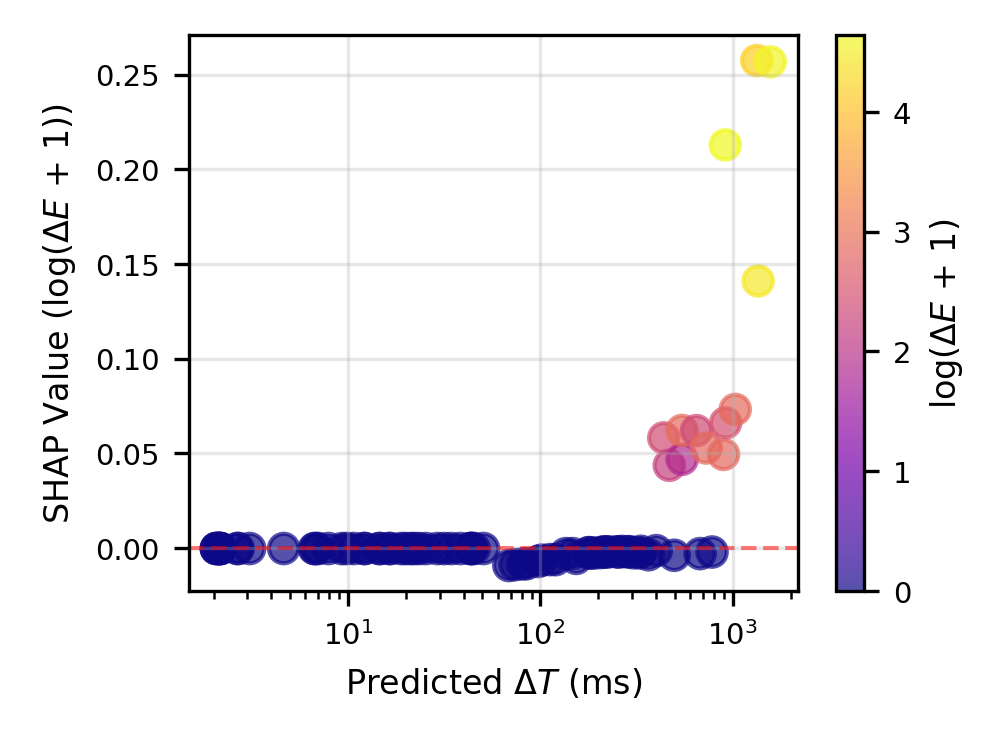}
    \caption{SHAP value for log($\Delta$E+1) versus predicted $\Delta \, T$ (ms), colored by
log($\Delta E$ + 1). }
    \label{fig:shapEnergy}
\end{figure}

\section*{Declaration of Generative AI and AI-assisted technologies in the writing process}
During the preparation of this work, the authors used Microsoft Copilot to review and enhance the clarity and grammar of the manuscript. After using this tool/service, the author(s) reviewed and edited the content as needed and take full responsibility for the content of the publication.

\bibliographystyle{elsarticle-num} 
\bibliography{References}
\end{document}